# Light emission from self-assembled and laser-crystallized chalcogenide metasurface

*Feifan Wang*[1,2,†], *Zi Wang*[1,†], *Dun Mao*[1,†], *Mingkun Chen*[1], *Qiu Li*[3,1], *Thomas Kananen*[1], *Dustin Fang*[1], *Anishkumar Soman*[1], *Xiaoyong Hu*[2], *Craig B. Arnold*[4]*, *Tingyi Gu*[1]*,

1 Electrical and Computer Engineering, University of Delaware, Newark, DE 19716, United States

2 State Key Laboratory for Mesoscopic Physics & Department of Physics Collaborative Innovation Center of Quantum Matter, Peking University, Beijing, 100871, P. R. China

3 Tianjin Key Laboratory of High Speed Cutting and Precision Machining, School of Mechanical Engineering, Tianjin University of Technology and Education, Tianjin 300222, P. R. China

4 Princeton Institute for the Science and Technology of Materials, Princeton University, Princeton, New Jersey 08544, United States

Subwavelength periodic confinement can collectively and selectively enhance local light intensity and enable control over the photo-induced phase transformations at the nanometer scale. Standard nanofabrication process can result in geometrical and compositional inhomogeneities in optical phase change materials, especially chalcogenides, as those materials exhibit poor chemical and thermal stability. Here we demonstrate the self-assembled planar chalcogenide nanostructured array with resonance enhanced light emission to create an all-dielectric optical metasurface, by taking advantage of the fluid properties associated with solution processed films. A patterned silicon membrane serves as a template for shaping the chalcogenide metasurface structure. Solution-processed arsenic sulfide metasurface structures are self-assembled in the suspended 250 nm silicon membrane templates. The periodic nanostructure dramatically manifests the local light-matter interaction such as absorption of incident photons, Raman emission, and photoluminescence. Also, the thermal distribution is modified by the boundaries and thus the photo-thermal crystallization process, leading to the formation of anisotropic nano-emitters within the field enhancement area. This hybrid structure shows wavelength selective anisotropic photoluminescence, which is a characteristic behavior of the collective response of the resonant guided modes in a periodic nanostructure. The resonance enhanced Purcell effect could manifest the quantum efficiency of localized light emission.

## 1. Introduction

Periodic perturbation of refractive index in subwavelength dimensions can boost efficiency of light emission device. Many nanophotonic structures, such as photonic crystals, plasmonic structures, gratings and resonators have been demonstrated to improve the light emission efficiency through enhancing light out-coupling efficiency [1,2] and spontaneous emission rate.[3-5] Nanophotonic engineering can lead to narrowband and directive light emission.[6-8] Chalcogenide materials with unique phase change properties have been explored for tunable thermal mission [9] or reflection [10] in infrared wavelength ranges. Photonic crystal (PhC) waveguide has been demonstrated on thin film chalcogenide by e-beam lithography, showing slow light and resonance enhanced parametric nonlinear process in material.[11,12] To reduce cost and improve scalability, solution processed chalcogenide nanostructure is recently reported through solution process and self-assembling.[13] In this work, we demonstrate the first active chalcogenide metasurface fabricated by self-assembling with tunable dimensions. The nanophotonic structure is tailored to enhance the light emission efficiency through resonance enhanced absorption of excitation, suppressing guided mode at emission wavelength and Purcell enhancement.[14-17] The two-dimensional hexagonal chalcogenide nanorod arrays are self-assembled [13] on a silicon template through solution processing.[18-24] Silicon nanophotonic structure with a few nanometer surface roughness enables delamination of top and bottom bulk chalcogenides during solvent evaporation leaving only a filled nanorod array. Nanophotonic confinement can simultaneously manifest local photon density for enhanced resonance absorption [25] and nonlinear process.[26] The guided mode with a concentration in the light emitting materials can significantly enhance the quantum efficiency of both stimulated Raman emission and spontaneous PL process, which plays a major role in the device compared to the leaky mode enhancement of light out-coupling efficiency.[27-29]

## 2. Localized photon and heat distribution in the hybrid metasurface

Fabrication of the light-emitting device is composed of two steps 1) assembling the $As_2S_3$ nanostructures in the silicon template; 2) local laser annealing for enhanced PL. The first step is for shaping the chalcogenide nanostructures and the second step removes the solvent related atomic defects and induces phase transitions in chalcogenide for effective light emission.

### 2.1. Self-assembled chalcogenide metasurface

Here we demonstrate a simple and scalable way of producing chalcogenide nanostructures, combining advanced silicon nanofabrication and solution processing. **Figure 1** shows the fabrication process and scanning electron microscope (SEM) images of a self-assembled chalcogenide-silicon metasurface structures. The silicon patterns (triangular lattice) are fabricated on a 250 nm thick silicon-on-insulator device layer via ultraviolet lithography and etching for reduced disorder scattering [30] (Figure S1). The lattice constant of the periodic structure is 415 nm, and we vary the hole radii from 80 to 150 nm with 10 nm per step (Figure 1b). The supporting silicon oxide buffer layer is removed via buffered oxide wet etching. $As_2S_3$ solution is prepared by uniformly dissolving the powder into n-propylamine solvent according to recipes in the prior literature.[14,18] The solution is then drop-casted onto the suspended silicon membrane, forming a thick $As_2S_3$ film on top of the Si layer. The sample is kept in the oxygen-free glovebox environment for several days at room temperature to evaporate out excess solvent. $As_2S_3$ dissolved in alkyl amines can be described as a nanocolloidal solution consisting of flat clusters, that internally retain the structure of the layer-like starting material but capped by ionic pairs of sulfide dangling bonds and alkyl ammonium molecules.[31] Without annealing, those solvent related atomic defect states suppress light emission from $As_2S_3$. [14]

Due to the fluid nature of the drop casted materials, the solution will accumulate in the hollow region [13] with excess chalcogenide material above and below the silicon membrane. The sample is left in glove box at room temperature overnight. Upon evaporating the excess solvent, the dried bulk chalcogenide cleanly delaminates from the smooth, flat silicon surface (Figure S2). Our previous work shows that the room temperature dried chalcogenide nanorods still contains solvent related residue (C-H/N-H bond),[32] which can be removed by hot plate annealing (170°C) or laser annealing.[14] Energy-dispersive X-ray spectroscopy (EDX) mapping is used to verify the chalcogenide arrays in the silicon structure (Figure S3), and little residual on silicon top surface. The results show the formation of uniform $As_2S_3$ triangular arrays without solvent-specific composition. With limited residual from the solvent, the surface roughness of the delaminated bare silicon region is limited to be a few nm, as verified by atomic force microscope imaging (Figure 1c).

## 2.2. Localized laser annealing

As shown in the cross-sectional image in Figure 2(a), isolated $As_2S_3$ nanorod arrays are self-assembled in the plane of the suspended silicon template. Laser annealing is then achieved by focusing a 532nm continuous wave laser onto the planar chalcogenide nanostructure. The laser spot size is about ~1 µm. The intensity profile is superimposed onto the SEM image in Figure 2(b). An example of the photothermal heat distribution in the middle of nanostructured plane (marked as a dashed blue line in Figure 2(d)) is shown in Figure 2(c) at a hole radius of 140 nm, simulated by the combination of the Discontinuous Galerkin Time-Domain algorithm (for Maxwell equation) and finite difference method (for heat transport equation), assuming an environmental temperature of 300 K.[33] As the excitation laser power is set at $40\mu W/\mu m^2$, the photothermal effect heats the local area to 170°C. The laser Gaussian profile is dramatically modified by the nanostructure. The modified mode profile leads to localized heating in plane (Figure S4) and out of the plane (Figure S5). The cross-sectional view shows that the highest temperature is found on the interface between the metasurface and bulk chalcogenide substrate (Figure 2(d)). The constructive interference between incident and reflected light from metasurface-bulk substrate chalcogenide interface can result in the highest optical field on the interface. The area with highest photon density has the highest heat generation rate. Beyond phase transition temperature, extra influence from the crystallization might slightly modify the temperature distribution. The top view of the hot spot distribution on the interface between the metasurface and bulk substrate is shown in the Figure 2(e). The local temperature can be up to 285°C, which is beyond the glass transition temperature of 191.7°C.[14]. The local photothermal heating removes the amine content and reconstructs the atomic structure from $As_2S_3$ to $As_4S_4$. The atomic state is stabilized after 10-second laser exposure (Figure S6). $As_4S_4$ exhibits higher PL intensity in the visible band, which likely comes from the loosely bound Wannier-Mott exciton in the atomic network of $As_4S_4$.[14] Figure 2(f) compares the PL spectra of bulk chalcogenide before and after laser crystallization. Enhancement factors up to 35 (grey dotted curve) are observed near 760 nm. The PL spectrum gives a major PL peak at 703 nm (1.76 eV) with 98 nm bandwidth and a minor peak at 745 nm (1.66 eV) with 30 nm bandwidth. PL spectra under different wavelength excitation are distinguished from each other, showing the multi-component nature of the laser crystallized $As_4S_4$ (Figure S7).

**3. Metasurface enhanced light emission**

The nanophotonic structure tailors both Raman emission and PL. The Raman emission has wavelengths close to excitation, narrow bandwidth and is weakly influenced by defect states accelerated nonradiative recombination, and thus used for characterizing the metasurface enhancement of excitation light. PL is broadband response and its intensity strongly influenced by those nonradiative recombination. The PL spectra depend on the superposition of resonance enhanced mode at excitation and emission wavelength. By comparing the metasurface enhancement of both emissions, we can obtain insights about the light emission enhancement mechanisms.

### 3.1. Raman emission and excitation enhancement

The frequency offset of the Raman emission to the excitation laser wavelength fingerprints the unique phonon mode of the target material. Characterization of the Raman emission illustrates the mechanism of the enhancement of excitation light through coupling into the guided mode with a mode profile that overlaps well with the $As_2S_3$ region. Local excitation photon density enhancement in $As_2S_3$ is examed by microRaman spectrometry. **Figure 3**(a) shows the normalized Raman spectra of chalcogenide metasurface with increasing hole radii of 70nm, 120nm, and 130nm (with 0.5 offset for clarity). In the hybrid nanostructure, the Raman peak near 305 cm$^{-1}$ is the transverse acoustic mode in silicon. The Raman peaks at 345cm$^{-1}$ represent asymmetric stretching vibrational modes of the $AsS_{3/2}$ pyramids.[34-36] The relative Raman peak intensity ratio of the As-S bond and silicon increases with hole radius ($R$). The enhancement of the 532 nm photon excitation initially increases with $R$, and then decreases as the $R$ becomes larger than 140nm ($R/a$=0.337). At $R$ = 140nm, the incident photon has the highest overlap factor/enhancement factor with the chalcogenide area (Figure 3b-c), and thus, the highest Raman peak intensity. The insets of Figure 2(b) show the top (left) and cross-sectional (right) view of the photon distribution in the hybrid nanostructure at $R$ = 140 nm, simulated by finite-difference time-domain method (FDTD). The total enhancement factor of an exciting photon is defined as the integral of the local electric field enhancement ($F_{Ex}(\lambda_0,r)$) over the chalcogenide region: $\int_{As_2S_3} |F_{Ex}(\lambda,r)|^2 dr$, where $\lambda_0$ is the wavelength of the excitation laser and $r$ is a vector of position. The field enhancement factor is calculated by the integral $\int_{As_2S_3} |F_{Ex}(\lambda_0,r)|^2 \times |F_{Em-Raman}(\lambda,r)|^2 dr$, where $F_{Em-Raman}$ is the Raman emission enhancement factor. The enhancement of the Raman emission wavelength is similar to the excitation, as their wavelength is closely spaced. The trend of simulated Raman emission intensity versus hole dimension (grey circles linked by the dashed line in Figure 3(b)) aligns well with the experimental data (blue dots with error bars in the same figure).

**3.2. Photoluminescence from the nanostructured chalcogenide emitter**

*3.2.1. Modeling metasurface modified PL spectrum*

Different from narrow band ultrafast Raman emission (femtosecond scale lifetime), PL is a non-parametric and broadband quantum process (nanosecond scale lifetime).[37,38] The emission bandwidth usually is across a few photonic guided/leaky modes. Selective enhancement of emission only occurs at resonance enhanced wavelengths. The resonant mode can effectively enhance the local density of states (LDOS) $\rho$, and thus, the spontaneous emission rate for radiative recombination. The local spontaneous emission rate is defined as $\Gamma_r = \frac{2\pi}{\hbar}\langle\langle\vec{d}\cdot\vec{E}\rangle\rangle^2 \rho$, where $d$ is the electric dipole moment, $E$ is the local electric field, and $\rho$ is the density of the electromagnetic modes or LDOS.[21] LDOS $\rho(\lambda,r)$ can be obtained through three-dimensional FDTD simulation. The internal quantum efficiency of the PL is $\eta = \Gamma_r/(\Gamma_r + \Gamma_{nr})$, where $\Gamma_r$ and $\Gamma_{nr}$ are the radiative and non-radiative recombination rates. The carrier recombination time ($1/\Gamma_{nr}$) of the chalcogenide semiconductor nanomaterial is usually at the picosecond scale, which is three orders of magnitude smaller than the radiation recombination lifetime ($1/\Gamma_r$). Because the quantum efficiency of the amorphous material is usually small ($\Gamma_r \ll \Gamma_{nr}$), $\eta$ ($\sim\Gamma_r/\Gamma_{nr}$) linearly increases with $\Gamma_r$. The local enhancement factor ($F_{Em}(\lambda,r)$) of the electric field for the emitted photon is $\rho/\rho_0$. $\rho_0$ is the LDOS in bulk. The overall emission intensity from the arsenic sulfide emitter is given by equation (1), assuming the intensity of light is well below the saturation level.

$$\frac{I_{phc}}{I_{bulk}} = C \int_{As_4S_4} |F_{Ex}(\lambda_0,r)|^2 \times |F_{Em}(\lambda,r)|^2 \times EE(\lambda,r) dr \tag{1}$$

Where $C$ is a constant describing the dipole alignment to the guided modes ($0<C<1$); $F_{Em}$ and $F_{Ex}$ are the enhancement factor of the local electric field for emitted (from 600 nm to 800 nm) and pump/excitation (532 nm) light respectively; $\lambda$ and $\lambda_0$ are the wavelengths of emission and pump light. *EE(λ,r)* is the optical power extraction efficiency of the local dipole emission, depending on the quality factor matching of intrinsic and out coupling quality factors of the mode and wavelength detunings to its resonant wavelength.[2]

*3.2.2. Contributions from field enhancement effect and light outcoupling efficiency*

In the hybrid metasurface, it is important to differentiate and understand individual contributions from the local field enhancement effect ($FE_{ex}FE_{em}$) and extraction efficiency (*EE*). Here we compare the experimentally measured PL from the metasurface with increasing hole radius and FDTD calculations of outcoupling efficiency and field enhancement factors (Figure 4). The measured PL spectra of 7 devices with consecutively increasing hole radius are plotted in Figure 4(a) (solid red line), compared to the line shape of PL from bulk (same as the one in Figure 2(f)). The PL intensity is normalized to the Raman signal of the As-S-As at 325 cm$^{-1}$, to ensure the normalized PL is proportional to the quantum efficiency and excludes effect from sample uniformity and changing filling factors, etc. Comparing the PL from devices with increasing hole radii, we observe that the PL efficiency increases with hole radius, and the resonance enhanced modes (highlighted by dashed grey curves) are blue shifted. The experimentally measured values align well with the trend of field enhancement simulated by FDTD simulation. Figure 4(b) plots the radius dependent outcoupling efficiency within 15 degrees to the Gamma point, which is correspondent to the band diagrams (Figure S8). As the dipole is confined in the chalcogenide region, the increased hole radius shifts those resonance enhanced modes to longer wavelengths. The trend of the field enhancement factors is opposite, as plotted in Figure 4(c), which incorporates the mode overlap among excitation, emission, and $As_4S_4$ region (Figure S9). And thus, we conclude that the PL enhancement is dominated by the contribution from local field enhancement.

Among all the PL measurements, the maximum enhancement is achieved at a hole radius of 140 nm at the wavelength of 670 nm, where both excitation and emission resonance modes are excited, along with good mode overlap with the $As_4S_4$ region. We compare the FDTD calculated spectra and experimental measurement in Figure 5(a). The blue solid curve shows the experimental measurement of the chalcogenide metasurface, and the black solid curve shows the FDTD simulated spectrum. Major spectral features are resembled in the simulation, including major peaks near 670nm, 700nm, 800nm, and ripples around 600 nm. The bulk contribution from the chalcogenide under the metasurface plane is marked via the blue dashed curve. Comparing the metasurface PL (solid blue curve) and bulk PL (blue dashed curve), a 4x enhancement is observed with field enhancement in the nanostructure.

*3.2.3. Directional PL*

All the measurement and FDTD simulations are made near the Gamma point (Figure 4, 5(a)). Then we explore the polar angle dependence of the PL. We measure the polar angle dependent Raman and PL spectra. The PL is also observed to be highly directional. The Raman scattering of As-S-As vary little at increasing tilting angle, with little influence from the nanophotonic structure. The red stars and blue crosses in Figure 5(b) represent Raman scatterings from $As_4S_4$ and silicon, respectively. In contrast, the PL intensity rapidly decreases with the incident angle, which drops to less than 10% at 8° tilting/polar angle. The empty grey squares, circles, and dots in Figure 5(b) mark the relative intensity of the three peaks on the $As_4S_4$ PL spectra. The directional PL can be attributed to metasurface nanostructure. As a control sample, a uniform thin chalcogenide film on silicon is examed with exact laser crystallization and measurement procedures as for metasurface samples. Only a 5% variation of the PL intensity is observed within 8 degrees of sample rotation in tilting/polar angle.

## 4. Conclusion

In conclusion, we present analytical and experimental investigations of self-assembled and laser-crystallized chalcogenide nano-emitters embedded within silicon template. Near-bandgap illumination is used to initiate localized laser crystallization (high power) and to probe the Raman and PL (low power). We first quantify the enhancement of photon density at excitation wavelength through micro Raman emission. A model incorporating excitation enhancement, Purcell factor and extraction efficiency is compared to experimentally measured PL spectra. from the nanostructured $As_4S_4$ emitter. The relation between peak wavelength shift and the rod dimension indicate the PL enhancement is dominant by the Purcell effect. The overall PL enhancement is attributed to the spatial overlap between the photonic guided modes at excitation and emission wavelengths.

## 5. Experimental Section

*Device fabrication:* The Silicon device is manufactured in a CMOS foundry. Deep-ultraviolet lithography defines the template features on intrinsic silicon-on-insulator substrates, followed by reactive ion etching.[29] The dimension of the nanostructured area is 20µm by 200µm, which is much larger than the laser spot size. The finely grounded $As_2S_3$ dissolves in n-propylamine for highly concerntrated solutions (0.4g/ml). The high concerntration improves the viscosity and thus change the flow properties of the solution. The concerntrated solution is then drop casted onto a Silicon template and dried, forming the self-assembled of $As_2S_3$ in the periodic voids.

Scanning electron microscope (SEM) images are taken with a Quanta 200 FEG environmental SEM at 5keV in high vacuum mode. Electron diffraction x-ray equipped with SEM is then used for material composition analysis (Figure S3).

*Optical measurements:* We characterize the chalcogenide metasurface by Raman spectra and PL. Raman spectra were collected by coupling the light scattered from the sample to an inVia Raman spectrometer through a ×20 objective, with a ~1 µm² laser spot size and 0.2 numerical aperture (Renishaw). The excitation wavelength is set at 532 nm.

*Numerical simulation:* The simulation of excitation mode enhancement is carried out by forwarding design. A plane wave of 532nm light is normally incident onto the metasurface plane. The broadband emission mode profile is performed by inverse design. A dipole source covering the light emission bandwidth of chalcogenide is placed above the metasurface plane. The spectral response in chalcogenide region is collected and summarized at four random locations the source placed. The spatial resolution is set at 5nm. The material absorption and dispersion of the silicon and chalcogenide are separately measured and taken into consideration in simulation.

The extraction efficiency of the local dipole emission and the field enhancement factors in Eq. (1) are simulated using the commercial software Lumerical FDTD solutions. To calculate the power extraction efficiency near the Gamma point, a dipole is set in the central hole of the metasurface in the model. The power is collected by a point monitor right above the central hole. Then the power spectrum is normalized to the spectrum of the dipole source in the model. To calculate the field enhancement factors, firstly a Gaussian source of 532nm is set above the central hole of metasurface as the excitation light, and a monitor is set to get the field profile inside the 3D photonic nanostructure. Then in another model, several Gaussian sources are set in the holes as emitters to get the field distribution profile of emission light.

**Supporting Information**
Supporting Information is available from the Wiley Online Library or the author.

**Acknowledgments**
The authors acknowledge experimental support from Y. Li and T. Heinz, and discussion with A. Rodriguez. This was supported in part by the National Science Foundation through the MRSEC Center (Grant No. DMR-1420541) and the AFOSR Young Investigator Program (FA9550-18-1-0300). Z. Wang are partially supported by the NASA Early Career Faculty Award (80NSSC17K0526). F. Wang and X. Hu acknowledge support from the National Natural Science Foundation of China (Grant Nos. 61775003, 11734001). Qiu Li thanks the support from the National Natural Science Foundation of China (No. 11772227). F. Wang, Zi Wang, and D. Mao contributed equally to this work.

Received: ((will be filled in by the editorial staff))
Revised: ((will be filled in by the editorial staff))
Published online: ((will be filled in by the editorial staff))

**Figures**

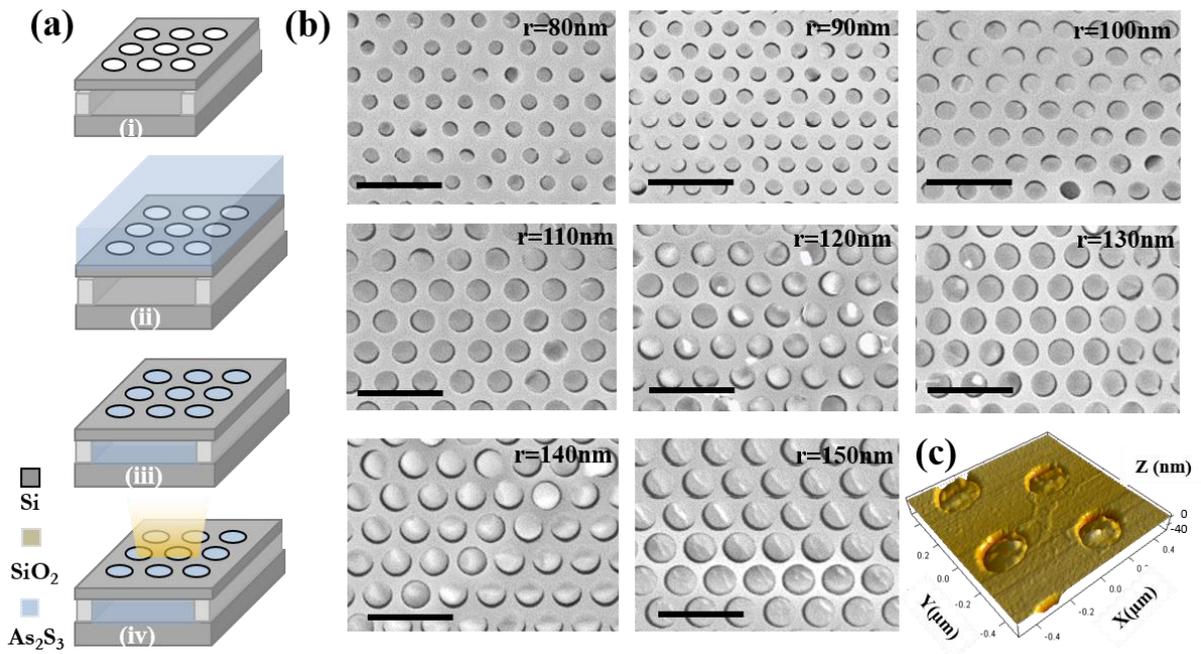

**Figure 1. Self-assembled chalcogenide metasurface with the solution process.** (a) Fabrication process. (i) Suspended planar silicon nanostructure. (ii) drop casting $As_2S_3$ solution. (iii) Reflow and Delamination. (iv) Top illuminated optical excitation. (b) Scanning electron microscope of delaminated $As_2S_3$ – silicon interface after step (iii). The silicon nanostructured template have hole radii from 80 nm to 150 nm, with 10 nm per step. Scale bars: 1µm. (c) Atomic force microscope image of the self-assembled surface topology.

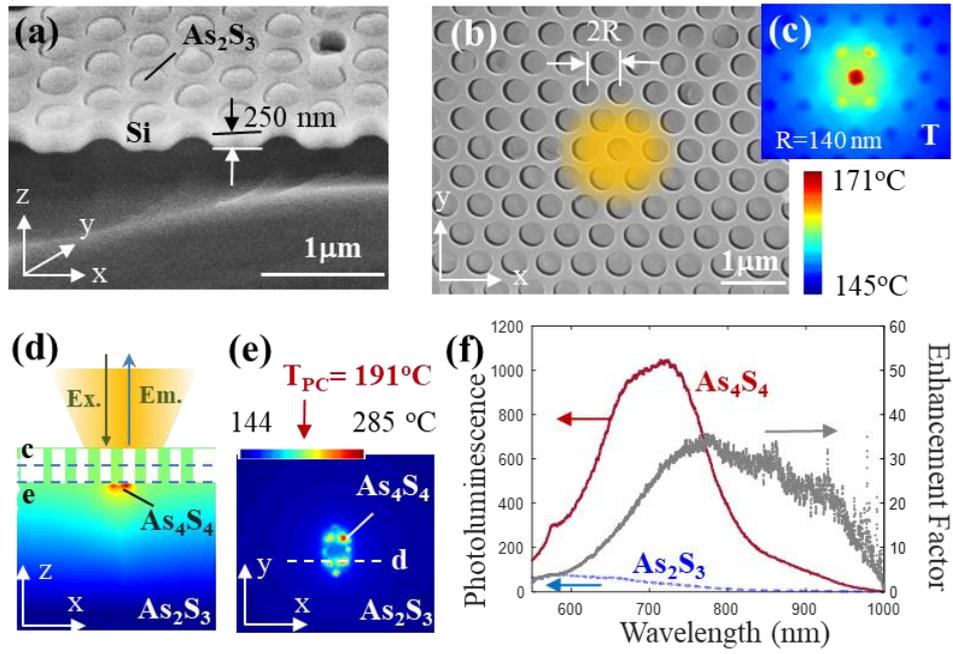

**Figure 2. Device geometry.** (a) Side and (b) top view of self-assembled chalcogenide metasurface in a silicon matrix. The yellow region indicates the intensity distribution of the top incident Gaussian beam profile, which is a continuous wave light centered at 532 nm. (c) Top view of the temperature distribution, with its z-axis marked in (d). (d) Temperature distribution on the cross-section, with its y-axis marked in (e). (e) Temperature distribution near the bottom plane of the hybrid metasurface, with its z-axis marked in (d). (f) Photoluminescence of $As_2S_3$ and laser crystallized $As_4S_4$.

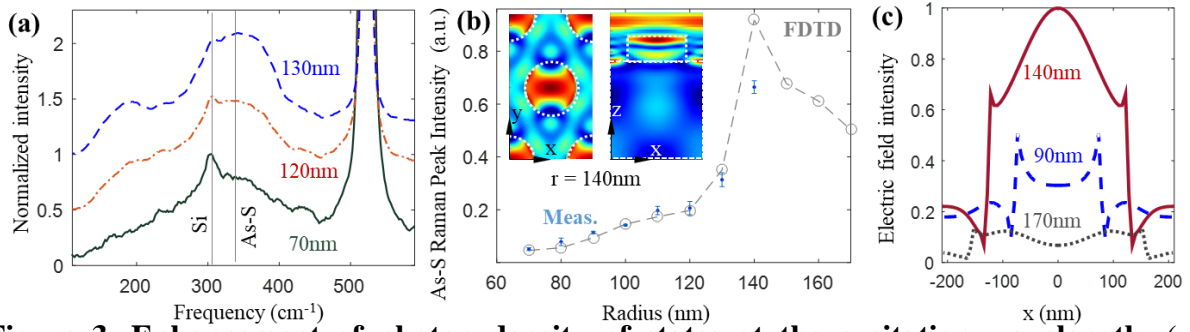

**Figure 3. Enhancement of photon density of states at the excitation wavelength.** (a) Raman spectrum of chalcogenide metasurface with a hole radius of 70nm, 120nm, and 130nm (accumulative vertical offset of 0.5 has been included in the spectra for clarity). (b) The intensity of Raman peak of chalcogenide (near 345cm$^{-1}$) versus the hole radius. Blue dots with error bar are experimental results, and the grey empty circles linked by the dashed line show theoretical prediction by FDTD calculations, incorporating the mode coupling effect in the chalcogenide region. Inset: Top view (left) and side view (right) of the electric field for an excitation wavelength of 532nm coupling into the metasurface plane, for the hole radius of 140nm. The chalcogenide region is highlighted by the dashed curves. (c) The electric field on the top surface of metasurface for hole radii of 90nm, 140nm, and 170nm.

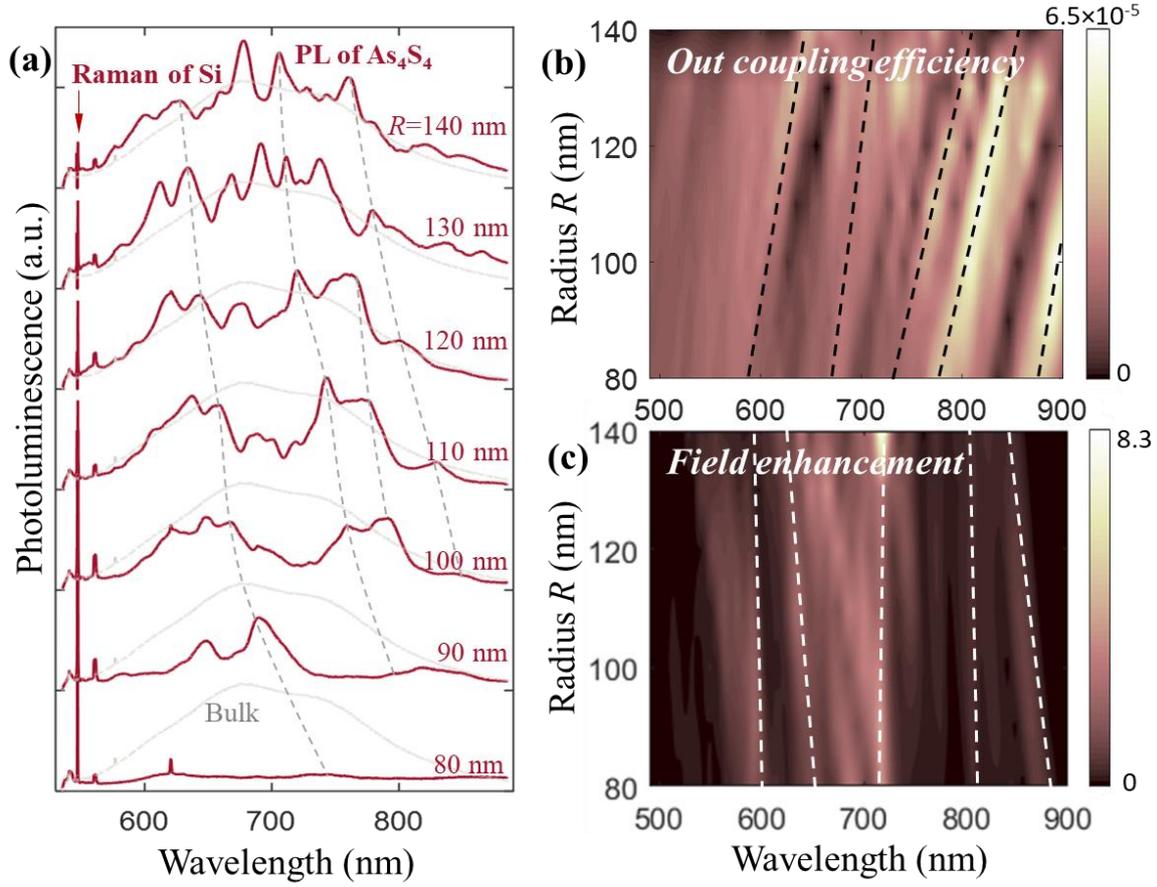

**Figure 4. Collective PL near Γ point.** (a) PL spectra of the metasurface emitters with hole radius from 80 nm to 140 nm in the step of 10nm. The grey dashed curve marks the shift of modes. The PL spectra from nanoemitter imbedded in metasurface (red solid curves) are compared to the line shape from the thin film (grey solid curves). (b) FDTD simulation of the optical power extraction efficiency of local dipole emission embedded in planar metasurface with hole radius as in (a). The trend of mode resonance shift with radius is highlighted in black dashed lines for guiding eyes. (c) The product of enhancement factors of lthe ocal electric field in nano-emitter: $\int_{As_4S_4} |F_{Ex}(\lambda_0, r)|^2 \times |F_{Em}(\lambda, r)|^2 dr$. The trend of geometry dependent modes is highlighted by dashed white lines.

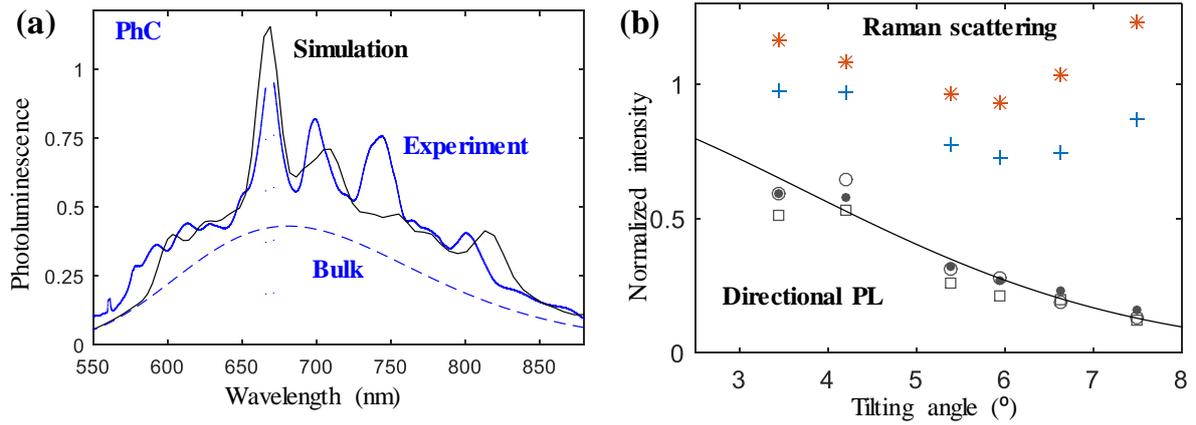

**Figure 5. Photoluminescence from the laser crystallized nano emitter.** (a) Photoluminescence spectrum of bulk chalcogenide (blue dashed curve), and metasurface before (blue solid curve) and after (black solid curve) laser irritation. Excitation laser of 532nm and 4mW is applied for probing the photoluminescence spectrums. (b) Angle-dependent Raman and photoluminescence emission intensity. The relative intensity to normal incidence for Raman peak of chalcogenide (blue cross), silicon (red star), and photoluminescence at three peaks (at a wavelength of 610nm, 650nm, and 725nm). The solid line is the fitting of Gaussian beam profile with the angular spreading of 14.9 º

**The table of contents entry:** Uniform chalcogenide metasurface is self-assembled on large scale silicon templates through solution process. The reducing viscosity allows transition from reflow to dewetting process with drying solvent. The subwavelength structured chalcogenide allows localized laser annealing with reduced spot size. The guided modes enhanced excitation and emission light manifest and photoluminescence signal from the localized emitter.

Keyword: Metasurface, optical nanostructures, laser processing, Raman, optical antennas

Feifan Wang, Zi Wang, Dun Mao, Mingkun Chen, Qiu Li, Thomas Kananen, Dustin Fang, Anishkuma Soman, Xiaoyong Hu, Craig Arnold*, Tingyi Gu*

**Title:** Light emission from self-assembled and laser-crystallized chalcogenide metasurface

ToC figure

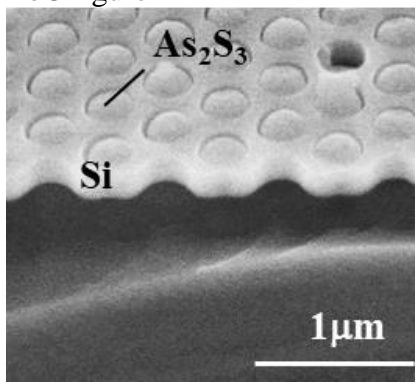

Supporting Information

**Light emission from self-assembled and laser-crystallized chalcogenide metasurface**

*Feifan Wang†, Zi Wang†, Dun Mao†, Mingkun Chen, Qiu Li, Thomas Kananen, Dustin Fang, Anishkumar Soman, Xiaoyong Hu, Craig B. Arnold*, Tingyi Gu**

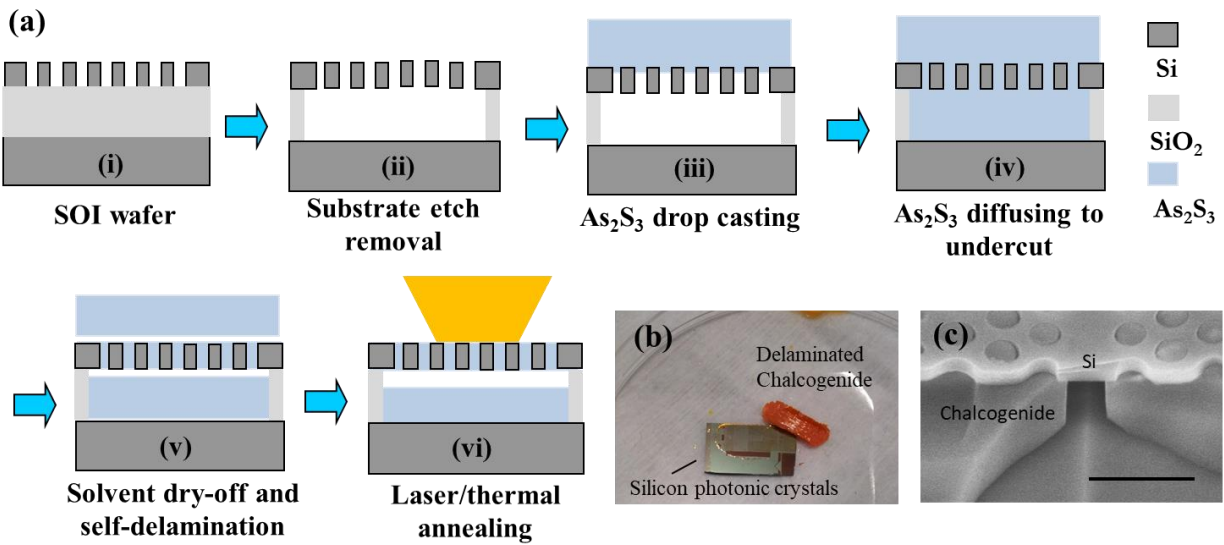

**Figure S1. Flow chart of device fabrication: solution process on the molded substrate.** (i) The substrate silicon patterns are fabricated on a 250 nm thick silicon-on-insulator device layer (ii) substrate removal by wet-etching (iii) $As_2S_3$ solution drop casting (iv) Solution diffusion to the undercut (v) Solvent dry-off and self-dewetting/delamination of the top layer (vi) Laser annealing for crystallization (b) optical image of the delaminated chalcogenide and silicon substrate (c) SEM image for the side view of the microstructures. Scale bar: 1µm.

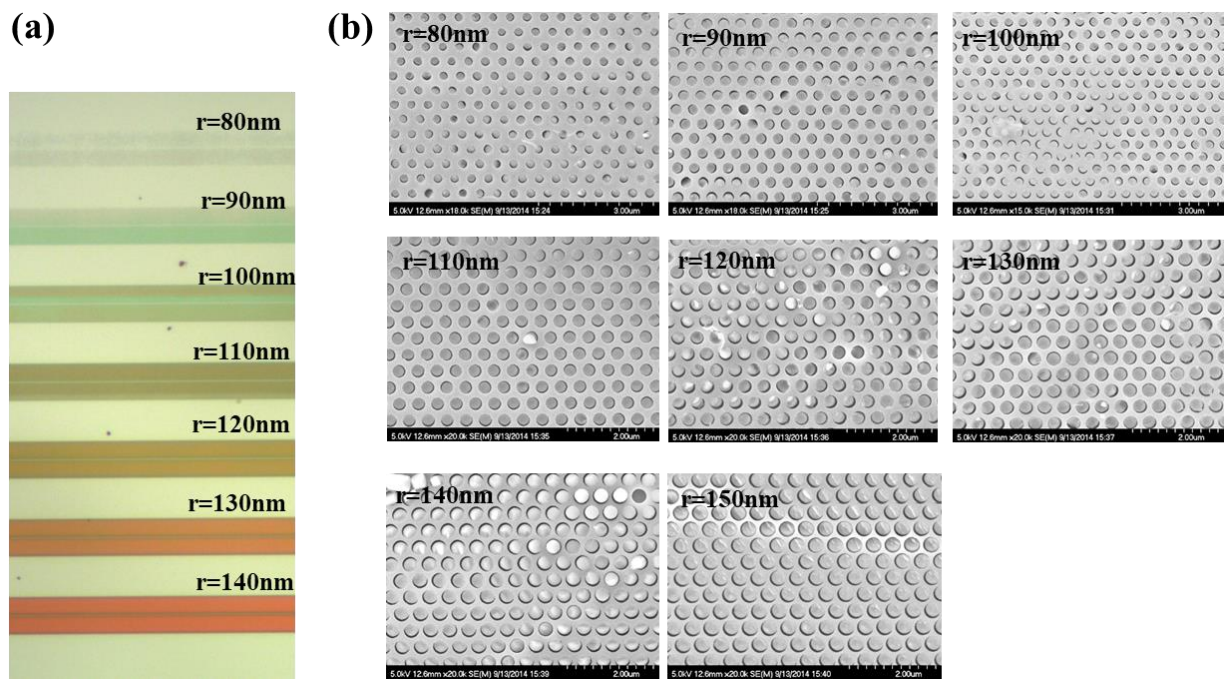

**Figure S2.** Optical (a) and SEM (b) image of the silicon nanostructures with varying geometric parameters.

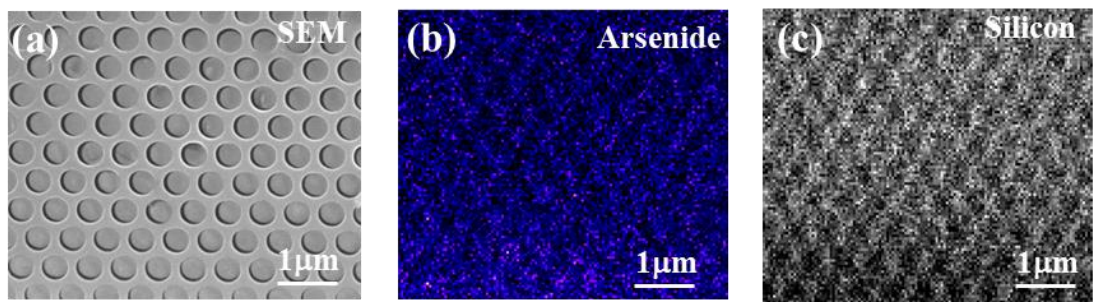

**Figure S3. Characterization of Chalcogenide-silicon metasurface.** (a) SEM of the device top view. (b) Energy-dispersive X-ray spectroscopy mapping of Arsenide element and (c) silicon in the hybrid metasurface structure.

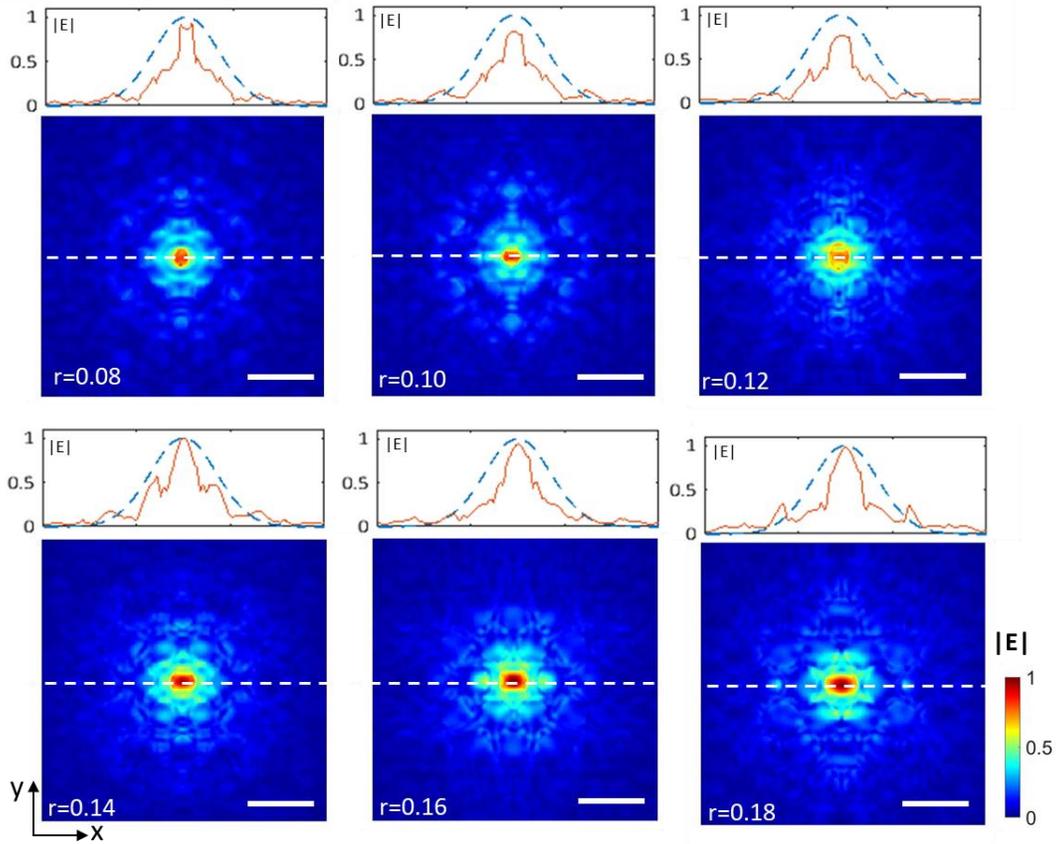

**Figure S4. Optical mode distribution in hybrid metasurface at an excitation wavelength of 532 nm** (XY plane at PhC-As$_2$S$_3$ interface). Scale bar: 1 µm.

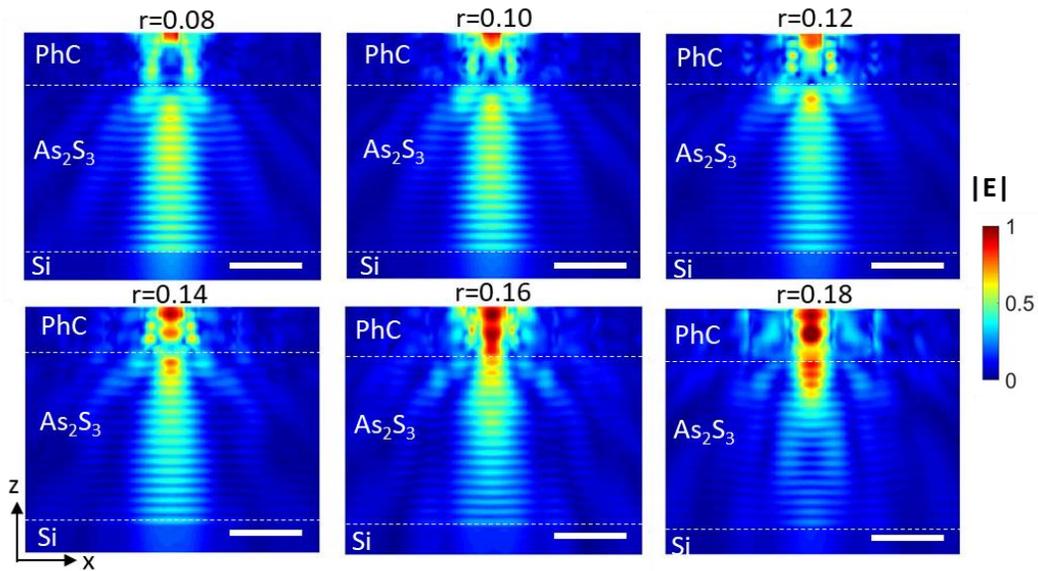

**Figure S5. Optical mode distribution in planar hybrid metasurface with various radius** (YZ plane of the dashed line in Figure S4).

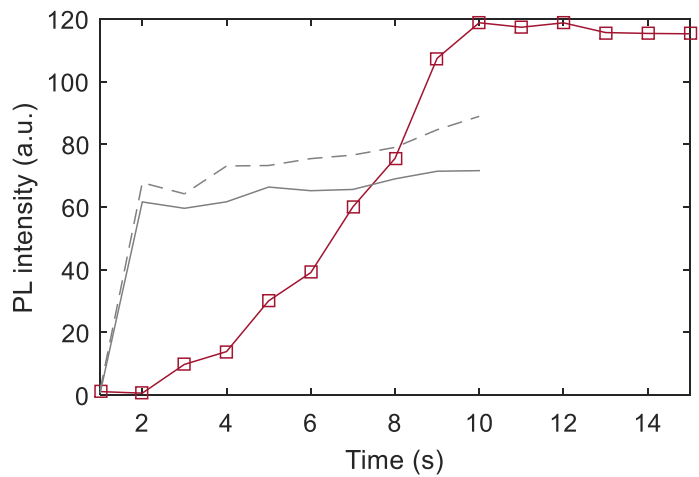

**Figure S6. Photoluminescence intensity versus the laser annealing time.** The grey solid/dash lines: the lower light intensity of 40 µW/µm². Red lines: 160 µW/µm².

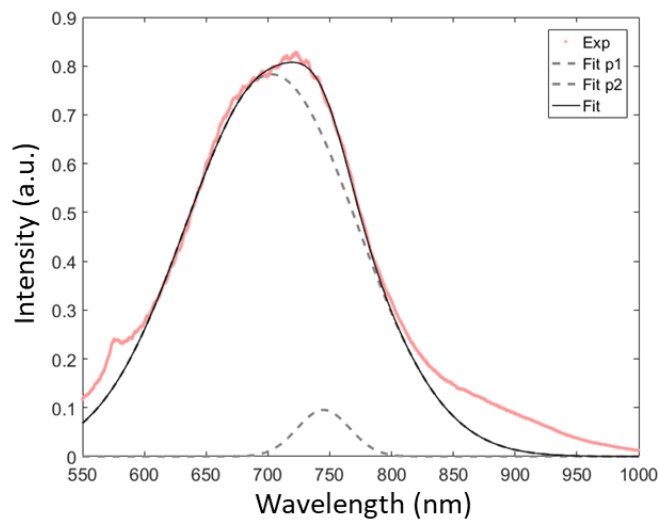

**Figure S7. Photoluminescence spectrum of laser crystallized As₄S₄.** Red dotted line: experimental data (the red curve in Figure 2f). Black solid curve: dual components gaussian fit. Grey dashed curve: two components in the Gaussian fitting curve.

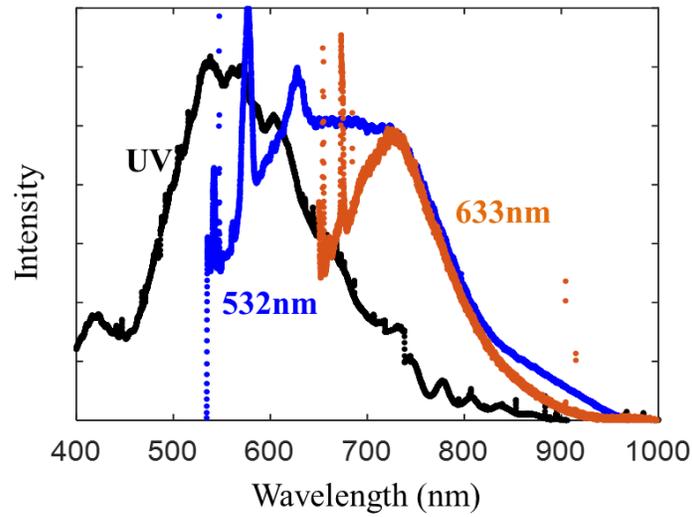

**Figure S8. Photoluminescence spectrum of laser crystallized thin film chalcogenide glass** with excitation laser of 355 nm (black), 532 nm (blue) and 633 nm (yellow).

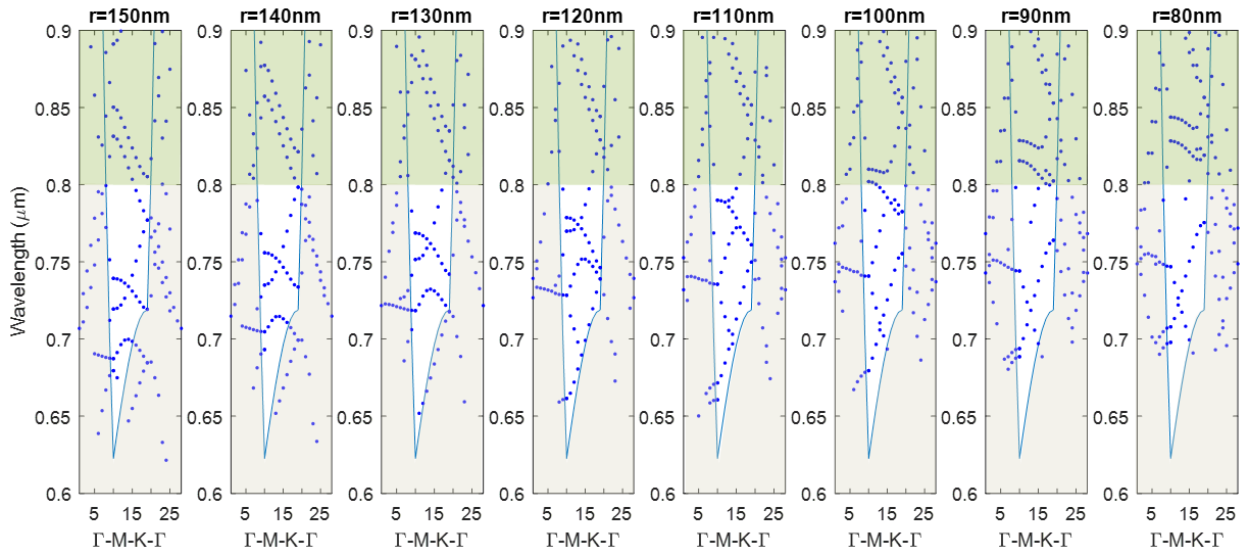

**Figure S9. Dispersion relation of chalcogenide-silicon metasurface slab for calculating the output efficiency.** The grey shaded area is the light cone, where each dot represents a state that can radiate in the air (leaky mode). The material light emission intensity is weak beyond 0.8 µm (green shaded), and the states at longer wavelength do not impact the line shapes of photoluminescence spectra.

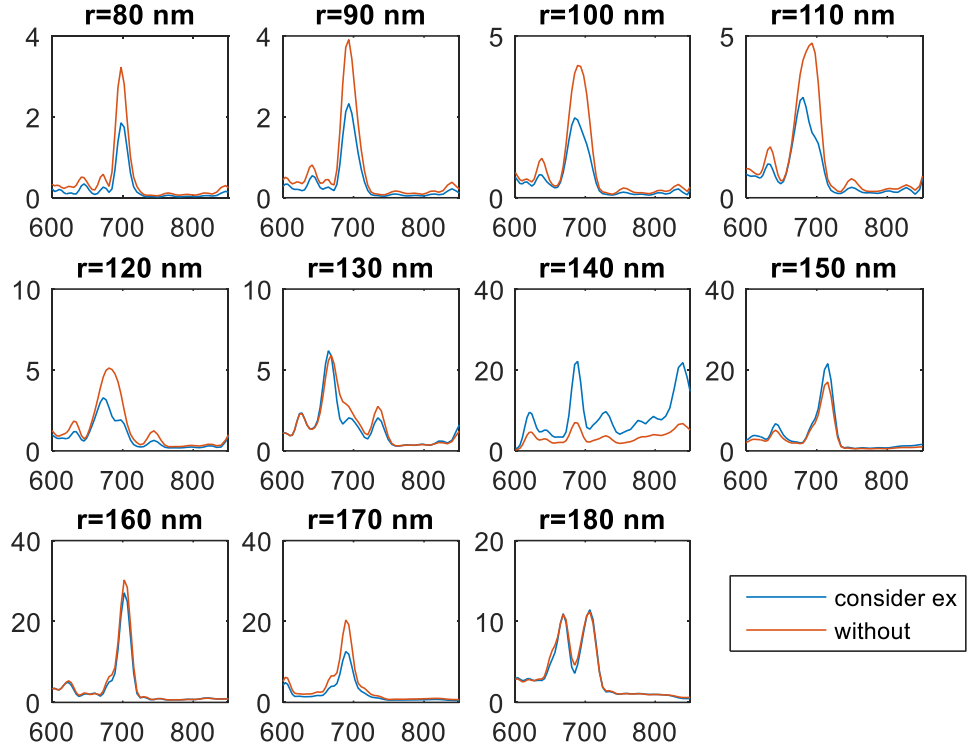

**Figure S10. Calculated enhancement factor spectra from the hybrid nanostructure with and without considering the optical mode overlapping with excitation.** Red curves: enhancement factor $\int_{As_4S_4} |F_{Em}(\lambda,r)|^2 dr$ ; Blue curves: enhancement factor $\int_{As_4S_4} |F_{Ex}(\lambda_0,r)|^2 \times |F_{Em}(\lambda,r)|^2 dr$.